\documentclass[aps,preprint]{revtex4}%
\usepackage{amsfonts}
\usepackage{amsmath}
\usepackage{amssymb}
\usepackage{graphicx}%
\ifx\pdfoutput\relax\let\pdfoutput=\undefined\fi
\newcount\msipdfoutput
\ifx\pdfoutput\undefined\else
\ifcase\pdfoutput\else
\ifx\paperwidth\undefined\else
\ifdim\paperheight=0pt\relax\else\pdfpageheight\paperheight\fi
\ifdim\paperwidth=0pt\relax\else\pdfpagewidth\paperwidth\fi
\fi\fi\fi
\begin{document}
\preprint{Manuscript}
\title{Spin magnetosonic shocks in quantum plasmas}
\author{A. P. Misra}
\email{apmisra@visva-bharati.ac.in}
\author{N. K. Ghosh}
\affiliation{Department of Mathematics, Visva-Bharati University, Santiniketan-731 235, India.}

\begin{abstract}
The one-dimensional shock structures of magnetosonic waves (MSWs) propagating
in a dissipative quantum plasma medium is studied. A quantum
magnetohydrodynamic (QMHD) model is used to take into account the quantum
force term due to Bohm potential and the pressure-like spin force term for
electrons. The nonlinear evolution (Korteweg de-Vries-Burger ) equation,
derived to describe the dynamics of small amplitude MSWs, where the
dissipation is provided by the plasma resistivity, is solved numerically to
obtain both oscillatory and monotonic shock structures. The shock strength
decreases with increasing the effects of collective tunneling and increases
with increasing the effects of spin alignment. The theoretical results could
be of importance for astrophysical (e.g., magnetars) as well as for ultracold
laboratory plasmas (e.g., Rydberg plasmas).



\end{abstract}
\eid{ }
\date{09.06.2008}
\startpage{1}
\endpage{102}
\maketitle



The subject of quantum plasmas have received a great attention in
investigating various collective quantum effects in plasmas [see e.g. Refs.
1-14]. The collective motion of Fermi particles in a magnetic field thus gives
rise a natural extension to the classical theory of magnetohydrodynamics (MHD)
in terms of the well-known quantum magnetoplasmas, which have potential
applications in astrophysical plasmas, such as pulsar magnetospheres,
magnetars. Moreover, the motion of particles with spin properties become
important in strong magnetic fields as a probe of quantum physical phenomena
[15-17] in the laboratory plasmas. Many of these studies are motivated on a
single particle properties. It is thus expected that the collective spin
effects can influence the propagation characteristics of waves in a strongly
magnetized quantum plasma [18-20]. Moreover, recent progress in producing
Rydberg plasmas may give rise to an interesting experimental evidence for the
dynamics of quantum plasmas. However, in such magnetized plasmas the thermal
energy of the particles can be very small compared to the typical Zeeman
energy of the particles. Recent investigations indicate that the spin
properties of the electrons and positrons can lead to interesting collective
effects in quantum magnetoplasmas [19]. More recently, it has been shown that
the electron spin 1/2 effect significantly modifies the dynamics [21] and
modulational instability domain [22] of magnetosonic solitary waves and the
collective effects in strongly magnetized plasmas [23].

There has also been much interests in investigating strucures and dynamics of
shock waves in various quantum plasma media [24-27]. The dynamics of classical
shocks is governed by a Korteweg de-Vries-Burger (KdVB) equation. A stationary
solution of the latter can be represented as an oscillatory shock. However,
when the dissipation overwhelms the dispersion and when the dissipative effect
is in balance with the nonlinearity, we indeed have the possibility of
monotonic shock waves. Unlike the classical fluids, quantum plasmas typically
exhibit dispersion due to the collective tunneling associated with the Bohm
potential instead of dissipation. For this reason, even a quantum shock
propagating with constant velocity in a uniform medium does not exhibit a
stationary structure. Transition from initial to compressed quantum media
occurs in the form of a train of solitons propagating with different
velocities and with different amplitudes.

In this letter, we derive a governing equatuion that describes the dynamics of
magnetosonic waves (MSWs) in a quantum electron-ion plasma. The governing KdVB
equation contains both dispersive term due to Bohm potential and the
dissipative term due to plasma resistivity (neglecting other effects viz.,
thermal conduction, viscosity etc.), and also the pressure-like spin quantum
force. Still, when the normalized zeeman energy$\sim1$ and the plasma
resistivity is small, we can recover monotonic transition of the oscillatory
shocks. The stationary shock solutions exist for the Mach number$\gtrsim15.$
The effects of collective tunneling and spin alignment influence the strength
of the shocks.

The basic set of equations governing the dynamics of the magnetosonic waves in
a quantum plasma reads [21,22]
\begin{equation}
\frac{\partial\rho}{\partial t}+\frac{\partial}{\partial x}\left(  \rho
v\right)  =0, \label{eqn1}%
\end{equation}%
\[
\rho\left(  \frac{\partial v}{\partial t}+v\frac{\partial v}{\partial
x}\right)  =-B\frac{\partial B}{\partial x}-c_{s}^{2}\rho\frac{\partial
}{\partial x}(\ln\rho)+
\]%
\begin{equation}
\beta\rho\frac{\partial}{\partial x}\left(  \frac{1}{\sqrt{\rho}}%
\frac{\partial^{2}\sqrt{\rho}}{\partial x^{2}}\right)  +\frac{\varepsilon
}{v_{B}^{2}}\rho\frac{\partial}{\partial x}\left[  \rho B\tanh(\varepsilon
B)\right]  , \label{eqn2}%
\end{equation}%
\begin{equation}
\frac{\partial B}{\partial t}+\frac{\partial}{\partial x}\left(  Bv\right)
-\gamma\frac{\partial^{2}B}{\partial x^{2}}=0, \label{eqn3}%
\end{equation}
where $\mathbf{B}$ is the magnetic field along the $z$-axis, i.e.
$\mathbf{B}=B(x,t)\hat{z}$, normalized by its equilibrium value $B_{0}$;
$\rho\equiv(m_{e}n_{e}+m_{i}n_{i})=\rho(x,t)$ is the total mass density
normalized by its equilibrium value $\rho_{0}$; and $\mathbf{v}\equiv
(m_{e}n_{e}\mathbf{v}_{e}+m_{i}n_{i}\mathbf{v}_{i})/\rho=v(x,t)\hat{x}$ is the
center of mass fluid velocity normalized by the Alfv{\'{e}}n speed
$C_{A}=\sqrt{B_{0}^{2}/\mu_{0}\rho_{0}}$. The space and time variables are
normalized by, respectively, $C_{A}/\omega_{ci}$ and the ion gyroperiod
$\omega_{ci}^{-1}\equiv(eB_{0}/m_{i})^{-1}$. Here $n_{e}(n_{i})$ is the
electron (ion) number density, $m_{e}(m_{i})$ is the electron (ion) mass,
$\mathbf{v}_{e}(\mathbf{v}_{i})$ is the electron (ion) fluid velocity and $e$
is the magnitude of the electron charge. Also, $\beta=2c^{2}(m_{e}%
/m_{i})\omega_{ci}^{2}\lambda_{C}^{2}/C_{A}^{4}$ , where $\lambda_{C}%
=c/\omega_{C}=\hbar/2m_{e}c$ is the Compton wavelength, $\omega_{C}$ is the
Compton frequency, $c$ is the speed of light in vacuum, $\hbar$ is the
Planck's constant divided by$2\pi$, $c_{s}=\sqrt{k_{B}(T_{e}+T_{i})/mi}$ is
the sound speed, where $T_{e}(T_{i})$ is the electron (ion) temperature, and
$k_{B}$ is the Boltzmann constant. Moreover, $\gamma=\eta\omega_{ci}/\mu
_{0}C_{A}^{2}$, where $\eta$ is the resistivity, $v_{B}^{2}=k_{B}T_{e}%
/m_{i}C_{A}^{2}=(1/\varepsilon)\mu_{B}B_{0}/m_{i}C_{A}^{2}$ with $\mu
_{B}=e\hbar/2m_{e}$ is the Bohr magneton and $\varepsilon=\mu_{B}B_{0}%
/k_{B}T_{e}$ is the temperature normalized Zeeman energy. Note that in Eq.(2)
[the second term on the right hand side] we have used the isothermal equation
of state for electrons as $P_{e}=k_{B}n_{e}T_{e}$ for one-dimensional
magnetosonic wave propagation across $B_{0}.$ One can also use the equation of
state for electrons as $P_{e}=m_{e}V_{Fe}^{2}n_{e}^{3}/3n_{0}^{2}$ for
one-dimensional propagation [3] or $P_{e}=m_{e}V_{Fe}^{2}n_{e}^{5/3}%
/5n_{0}^{2/3}$ in three-dimension, assuming a local zero-temperature Fermi
distribution [28]. The last two terms in the right-hand side of Eq.(3) are due
to the effects of collective tunneling and spin alignment, respectively, and
in Eq.(3) we have neglected the inertial term.

In order to investigate the dynamics of MSWs, we employ the standard reductive
perturbation technique (RPT) with the following stretching
\begin{equation}
\xi=\epsilon^{1/2}(x-v_{0}t),\tau=\epsilon^{3/2}t, \label{eqn4}%
\end{equation}
where $\epsilon$ is a small expansion parameter and $v_{0}$ is the wave phase
velocity normalized by $C_{A}$. The dynamical variables are expanded as
\[
\rho=1+\epsilon\rho_{1}+\epsilon^{3/2}\rho_{2}+\epsilon^{2}\rho_{3}+...,
\]%
\begin{equation}
v=\epsilon v_{1}+\epsilon^{3/2}v_{2}+\epsilon^{2}v_{3}+..., \label{eqn5}%
\end{equation}%
\[
B=1+\epsilon B_{1}+\epsilon^{3/2}B_{2}+\epsilon^{2}B_{3}+....
\]

Now, substituting the expressions [Eqs.(5)] into the Eqs. (1)-(3) and
collecting the terms in different powers of $\epsilon$ we obtain in the lowest
order of $\epsilon$
\begin{equation}
\rho_{1}=B_{1},\text{ }v_{1}=v_{0}(B_{1}-1), \label{eqn6}%
\end{equation}
together with the linear dispersion relation:
\begin{equation}
v_{0}=\sqrt{1+c_{s}^{2}-\frac{\varepsilon}{v_{B}^{2}}\left(  2\tanh
\varepsilon-\varepsilon\sec\text{h}^{2}\varepsilon\right)  }. \label{eqn7}%
\end{equation}

From the next order of $\epsilon,$ we obtain%

\begin{equation}
\rho_{2}=B_{2}+\frac{\gamma}{v_{0}}\frac{\partial B_{1}}{\partial\xi}%
,v_{2}=v_{0}B_{2}+\gamma\frac{\partial B_{1}}{\partial\xi}-v_{0} \label{eqn8}%
\end{equation}

and
\begin{equation}
v_{0}^{2}+v_{0}v_{2}=\left[  1-\frac{\varepsilon}{v_{B}^{2}}\left(
\tanh\varepsilon+\varepsilon\sec\text{h}^{2}\varepsilon\right)  \right]
B_{2}+\left(  c_{s}^{2}-\frac{\varepsilon}{v_{B}^{2}}\tanh\varepsilon\right)
\rho_{2} \label{eqn9}%
\end{equation}

Inserting Eq.(8) into the Eq.(9) we obtain%

\begin{equation}
\gamma\left(  v_{0}^{2}-c_{s}^{2}+\frac{\varepsilon}{v_{B}^{2}}\tanh
\varepsilon\right)  \frac{\partial B_{1}}{\partial\xi}=0 \label{eqn10}%
\end{equation}

Since the second factor in Eq.(10) is non-zero by means of Eq.(7) and also
$\partial B_{1}/\partial\xi\neq0,$ $\gamma$ should be at least of the first
order of $\epsilon$, so that $\gamma\partial B_{1}/\partial\xi$ becomes of the
order of $\epsilon^{2},$ and it will be included in the equations for the
order of $\epsilon^{2}.$ Collecting the terms in powers of $\epsilon^{2}$ and
eliminating the quantities $\rho_{3},v_{3}$ [the coefficient of $B_{3}$
becomes zero by Eq.(7)] we obtain with the help of Eq.(6) the required KdVB equation%

\begin{equation}
\frac{\partial b}{\partial\tau}+Pb\frac{\partial b}{\partial\xi}%
+Q\frac{\partial^{3}b}{\partial\xi^{3}}+R\frac{\partial^{2}b}{\partial\xi^{2}%
}=0, \label{eqn11}%
\end{equation}

where $b\equiv B_{1}$ and the coefficients $P,Q$ and $R$ are given by%

\begin{equation}
P=\frac{1}{2v_{0}}\left[  3-v_{0}^{2}+2c_{s}^{2}-\frac{\varepsilon}{v_{B}^{2}%
}\left(  8\tanh\varepsilon+7\varepsilon\sec\text{h}^{2}\varepsilon
-2\varepsilon^{2}\tanh\varepsilon\sec\text{h}^{2}\varepsilon\right)  \right]
, \label{eqn12}%
\end{equation}

\begin{equation}
Q=-\frac{\beta}{4v_{0}},\text{ }R=\frac{\gamma}{2v_{0}^{2}}\left(  c_{s}%
^{2}-M^{2}-\frac{\varepsilon}{v_{B}^{2}}\tanh\varepsilon\right)  .
\label{eqn13}%
\end{equation}

Note that the spin quantum effects are embedded in all of $P,Q$ and $R$,
whereas the dispersion due to quantum diffraction and dissipation due to
plasma resistivity are in $Q$ and $R$ respectively. We now numerically solve
the Eq.(11) directly in order to obtain nonstationary shock solutions. In the
numerical scheme the KdVB equation (11) is advanced in time with a standard
fourth-order Runge-Kutta scheme with a time step of $10^{-4}$ s. The spatial
derivatives are approximated with centered second-order difference
approximations with a spatial grid spacing of $0.2$ m. The profile of the
oscillatory shock solution of Eq.(11) for the parameter values $B_{0}=0.14$T,
$T_{e}=0.09$K, $\varepsilon=1.04,$ $n_{0}=10^{30}$m$^{-3},$ $\lambda\equiv
\eta/\mu_{0}=0.001$ is shown in Fig.1. The train of oscillations propagates
together with the shock with the same velocity. As the role of spin force
increases, the shock strength decreases and the oscillations ahead the shock
becomes less in number, in which the first few oscillations are very close to
the magnetosonic solitons. The oscillations decay quite slow as the role of
quantum diffraction increases. The plot with $\lambda=0.01$ shows the
monotonic transition (Fig.2) from the oscillatory shocks shown in Fig.1.
Increasing further the role of quantum effects, we can not observe the
oscillatory shock transition from the Fig.2.

The stationary solution of Eq.(11) can also be obtained by transforming to the
moving frame of reference $\zeta=\xi-V\tau=\sqrt{\epsilon}(\omega_{ci}%
/C_{A})\left[  x-C_{A}(v_{0}+\epsilon V)t\right]  .$ The KdVB equation then
reduces to the following system:%

\begin{equation}
\frac{db}{d\zeta}=a,\text{ }\frac{da}{d\zeta}=-\frac{1}{Q}\left(  Ra+\frac
{P}{2}b^{2}-Vb+V-\frac{P}{2}\right)  \label{eqn14}%
\end{equation}

The system of equations (14) has two singular points, namely $(1,0)$ and
$\left(  2V/P-1,0\right)  $ which are stable node or focus according as
$R^{2}+4VQ\lessgtr0$ (since $Q<0$). A stable focus corresponds to an
oscillatory shock (Fig.3) (dispersion dominant), while a stable node gives
rise monotonic shocks (Fig.4) (dissipation dominant). The shock strength is
given by%

\begin{equation}
\left[  \epsilon b\right]  _{\max}=\epsilon\left(  \frac{2V}{P}-1\right)
=\frac{2v_{0}}{P}(M-1)-\epsilon, \label{eqn15}%
\end{equation}
where we have defined the shock Mach number $M$ as the ratio of the velocity
$C_{A}(v_{0}+\epsilon V)$ of the nonlinear magnetosonic wave to the linear
wave velocity $C_{A}v_{0}$ by%

\begin{equation}
M=1+\epsilon\frac{V}{v_{0}}. \label{eqn16}%
\end{equation}

Numerical solutions of Eq.(14) are shown in Figs. 3 and 4 for the same
parameter values as in Figs.1 and 2 respectively, but for $V=20.$ We find that
for these sets of parameters shock solutions exist for $V\geq15.$ As the value
of $V$ increases, the number of oscillations ahead of the shock increases with
decreasing the shock amplitude near $\zeta=0.$ Also, as the value of the
zeeman energy $\varepsilon$ decreases, the shock strength increases and the
oscillations decay quite slowly forming long wave train, while for large value
of $\varepsilon,$ the oscillations decay quite fast. Numerical simulation also
reveals that the shock strength decreases with increasing the particle numer
density and decreasing the electron temperature, while it increases with the
strength of the ambient magnetic field.

To conclude, we have investigated the effects of quantum tunneling and spin
alignment on the magnetosonic shock structures in a dissipative quantum plasma
medium. The numerical solutions of the KdVB equation exhibit both stationary
and nonstationary oscillatory/ monotonic shock solutions in the quantum
regime. Such significant modifications of the shock structures in our quantum
plasma are completely a new feature relevant for astrophysical and ultracold
laboratory plasmas.

{\Large Acknowledgments}

This work was partially supported by the Special Assistance Program (SAP)
through UGC Sanction Letter no. F.510/8/DRS/2004 (SAP-I), Govt. of
India.\newpage

{\LARGE References}

{[1]} \ L.G. Garcia, F. Haas, L.P.L. de Oliveira, and J. Goedert, Phys. Plasmas

12 (2005) 12302.

{[2] } F. Haas, G. Manfredi, M.R. Feix, Phys. Rev. E 62 (2000) 2763; F.

Haas, G. Manfredi, J. Goedert, \textit{ibid} 64 (2001) 026413; Braz. J. Phys. 33

(2003) 128.

{[3] G. Manfredi, Fields Inst. Commn. }46{ (2005) 263}.

{[4] } F. Haas, L.G. Garcia, J. Goedert, G. Manfredi, Phys. Plasmas 10 (2003)

3858.

{[5] } B. Shokri, S.M. Khorashady, Pramana, J. Phys. 61 (2003) 1.

{[6]} \ B. Shokri, A. A. Rukhadze, Phys. Plasmas 6 (1999) 3450.

{[7] } P.K. Shukla, S. Ali, L. Stenflo, M. Marklund, Phys. Plasmas 13

(2006) 112111; M. Marklund, P.K. Shukla, Rev. Mod. Phys. 78 (2006) 591.

{[8] } P.K. Shukla , L Stenflo, Phys. Lett. A 357 (2006) 229; New J. Phys. 8

(2006) 111.

[9] \ P.K. Shukla, B. Eliasson, Phys. Rev. Lett. 96 (2006) 245001; 99 (2007)

096401.

[10] D. Shaikh, P.K. Shukla Phys. Rev. Lett. 99 (2007) 012502.

{[11]} A.P. Misra, A.R. Chowdhury, Phys. Plasmas 13 (2006) 072305.

{[12]} A.P. Misra, C. Bhowmik, Phys. Lett. A \ 369 (2007) 90; A.P. Misra,

D. Ghosh, A.R. Chowdhury, \textit{ibid.} 372 (2007) 1469.

{[13]} A.P. Misra, P.K. Shukla, Phys. Plasmas \ 14 (2007) 082312; C.

Bhowmik, A.P. Misra, P.K. Shukla; \textit{ibid}. 14 (2007) 122107; A.P.

Misra, C. Bhowmik, \textit{ibid.} 14 (2007) 012309.

[14] A.P. Misra, P.K. Shukla, C. Bhowmik, Phys. Plasmas 14 (2007) 082309;

A.P. Misra,\textit{\ ibid}. 14 (2007) 1.

[15] \ M.W. Walser, C.H. Keitel, J. Phys. B: At. Mol. Opt. Phys. 33 (2000)

L221.

[16] Z. Qian, Vignale, Phy. Rev. Lett. 88 (2002) 056404.

[17] R.L. Liboff, Europhys. Lett.\textbf{\ }68 (2004) 577.

[18] \ M. Marklund, G. Brodin, Phys. Rev. Lett. 98 (2007) 025001.

[19] G. Brodin, M. Marklund, New J. Phys. 9 (2007) 277.

[20] S.C. Cowley, R.M. Kulsrud, E. Valeo, Phys. Fluids 29 (1986) 430.

[21] M. Marklund, B. Eliasson, P.K. Shukla, Phy. Rev. E 76 (2007) 067401.

[22] A.P. Misra, P.K. Shukla, Phys. Plasmas 15 (2008) 052105.

[23] G. Brodin, M. Marklund, Phys Plasmas 14 (2007) 112107.

[24] K. Roy, A.P. Misra, P. Chatterjee, Phys. Plasmas 15 (2008) 032310.

[25] B. Sahu, R. Roychoudhury, Phys. Plasmas 14 (2007) 072310.

[26] M. Marklund, D.D. Tskhakaya, P.K. Shukla, Eur. Phys. Lett. 72 (2005)

950.

[27] V. Bychkov, M. Modestov, M. Marklund, arXiv:0801.1295v2

[physics.plasm-ph] 22 Feb. 2008.

[28] L.D. Landau, E.M. Lifshitz, Statistical Physics, part 1, 

Butterworth-Heinemann, Oxford, 1980.

\end{document}